\documentclass{article}
\usepackage{bm}
\usepackage{amsmath}
\begin{document}
\twocolumn[
\title{{\sc The Papapetrou equations\\and supplementary conditions}}
\author{O. B. Karpov\\
      {\it {\small Moscow State Mining University, Moscow 119991, Russia}}}
\date{}
\maketitle
\vspace{-8ex}%
\renewcommand{\abstractname}{ }\abstractname
   \begin{abstract}
    On the bases of the Papapetrou equations with various
supplementary conditions and other approaches a comparative
analysis of the equations of motion of rotating bodies in general
relativity is made. The motion of a body with vertical spin in a
circular orbit is considered. An expression for the spin-orbit
force in a post-Newtonian approximation is investigated. The
relativity of motions and of the fulfilment of the third Newton
law in the general relativistic two-body problem is discussed.
  \end{abstract}
  \vspace{5ex}%
  ]

    \section{Introduction}

    The only covariant general-relativistic equations of motion of
spinning test particles are well-known Papapetrou equations
\cite{1,2} reduced by Dixon \cite{3,4} to the form
\begin{equation}\label{eq:1}
\dot{P}^{\alpha}=-\frac{1}{2}R^{\alpha}{}_{\beta\mu\nu}u^{\beta}S^{\mu\nu},
\end{equation}
\begin{equation}\label{eq:2}
\dot{S}^{\alpha\beta}=2P^{[\alpha}u^{\beta ]}
\end{equation}
where $P^{\alpha}=\int T^{\alpha\beta}\,d^{3}\mathcal{s}_{\beta}$
is the 4-momentum of a body and $R_{\alpha\beta\mu\nu}$ is the
Riemann tensor of the background space-time. The antisymmetric
spin tensor with respect to an event $z^{\alpha}(s)$,
$\zeta^{\alpha}=x^{\alpha}-z^{\alpha}$,
\begin{equation}\label{eq:3}
S^{\alpha\beta} =
2\int\zeta^{[\alpha}T^{\beta]\gamma}\,d^{3}\mathcal{s}_{\gamma}
\end{equation}
depends on the representing world line determined by the tangent
4-vector
\[
u^{\alpha}=dz^{\alpha}/ds, \quad u^{\alpha}u_{\alpha}=1.
\]
The dot denotes a covariant derivative along $u^{\alpha}$,
$D/ds=u^{\alpha}\nabla_{\alpha}$.

    An essential feature of the Papapetrou equations (\ref{eq:1}) and
(\ref{eq:2}) is the freedom of a specific definition of the
representing world line $u^{\alpha}$. This freedom manifests
itself in the fact that the system (\ref{eq:1}) and (\ref{eq:2})
is not complete and the number of unknowns exceeds the number of
equations by three (the number of independent components
$u^{\alpha}$). Therefore the world line can be determined
arbitrarily from physical considerations. For example, one can
require the tangency of the 4-momentum to the world line of a
representing point \cite{5,6}
\begin{equation}\label{eq:4}
P^{\alpha} \propto u^{\alpha}.
\end{equation}
Then the spin tensor is parallel transported along $u^{\alpha}$
\begin{equation}\label{eq:5}
\dot{S}^{\alpha\beta}=0.
\end{equation}
Usually, to Eqs.(\ref{eq:1}) and (\ref{eq:2}), supplementary
conditions are added which single out the world line of the center
of mass (CM) as a representing path. The spin tensor with respect
to the CM, determined in a given reference frame with a tangent
vector of congruence $\tau^{\alpha}$, satisfies the condition
\begin{equation}\label{eq:6}
S^{\alpha\beta}\tau_{\beta}=0,
\end{equation}
which closes the system (\ref{eq:1}) and (\ref{eq:2}). In
stationary space-time it is natural to direct the $\tau^{\alpha}$
along the time-like Killing vector $\xi^{\alpha}$
\[
S^{\alpha\beta}\xi_{\beta}=0.  \eqno (6')
\]
Such a definition of the CM (\ref{eq:6}) will be called the
Corinaldesi supplementary  condition \cite{7}.

    The supplementary condition (\ref{eq:4}) also closes the system of the
Papapetrou equations. Then the equation
$S^{\alpha\beta}\tau_{\beta}=0$ entirely determines the frame
$\tau$, in which the $u^{\mu}$ is the CM 4-velocity. The parallel
transport (\ref{eq:5}) specifies this frame by the relation
$S^{\alpha\beta}\dot{\tau}_{\beta}=0$.

    For singling out the CM, determined in the rest frame of a body
(the intrinsic CM), it is necessary to direct the vector
$\tau^{\alpha}$ along the $P^{\alpha}$,
\begin{equation}\label{eq:7}
S^{\alpha\beta}P_{\beta}=0.
\end{equation}
This is the supplementary Dixon condition \cite{3,4}. Introducing
side by side with the kinematic 4-velocity $u^{\alpha}$, a dynamic
4-velocity $U^{\alpha}$
\begin{equation}\label{eq:8}
U^{\alpha}=P^{\alpha}/\sqrt{P^{\lambda}P_{\lambda}},
\end{equation}
one can assume the condition (\ref{eq:7}) to be a particular case
of Eq.(\ref{eq:6}) for $\tau^{\alpha}=U^{\alpha}$. In curved
space-time the intrinsic CM moves relative to the rest frame (see
Eq.(\ref{eq:49})). The CM in the frame in which it rests is
determined by the Pirani supplementary condition \cite{8}
$\tau^{\alpha}=u^{\alpha}$
\begin{equation}\label{eq:9}
S^{\alpha\beta}u_{\beta}=0.
\end{equation}
The Pirani CM can move in the rest frame $U$ even in flat
space-time (see Eq.(\ref{eq:23})).

    Until recently, the adequate choice of supplementary conditions
and the physical consequences of such a choice have been the
subject of wide discussion: unreserved use of the Pirani condition
\cite{9,10} and the solution of Eqs.(\ref{eq:1})--(\ref{eq:2}) in
the ultrarelativistic case \cite{11} when it greatly differs from
the Dixon condition; categorical rejection of a physical nature of
the Pirani condition \cite{3,4} and the use of the Dixon condition
\cite{12}; application of Corinaldesi condition even in
nonstationary space-time \cite{13}; the assertion \cite{5} about
the unphysical character of the Papapetrou equations in the case
of violation of the condition (\ref{eq:4}); and in Ref. \cite{6}
it is assumed that supplementary conditions act as external
nongravitational forces. In works \cite{14, 15} the CM freedom is
considered as unacceptable (shift of an electron CM is nonsense),
Papapetrou equations are supposed to be incorrect even in
post-Newtonian approximation and noncovariant description
differing essentially from the Eqs.(\ref{eq:1})--(\ref{eq:2}) is
constructed.

    In the present paper, we investigate the Papapetrou equations
and compare the conclusions to which different supplementary
conditions and alternative approaches lead. For comparison, we
introduce the following system of notation.

    The body mass: in the rest frame of a body,
\begin{equation}\label{eq:10}
M_{0}=P^{\alpha}U_{\alpha},
\end{equation}
in the frame $u$ comoving with the CM,
\begin{equation}\label{eq:11}
M=P^{\alpha}u_{\alpha}=M_{0}U^{\alpha}u_{\alpha},
\end{equation}
and in an arbitrary frame $\tau$,
\begin{equation}\label{eq:12}
m=P^{\alpha}\tau_{\alpha}=M_{0}U^{\alpha}\tau_{\alpha}=m_{0}u^{\alpha}\tau_{\alpha}.
\end{equation}

    In the frame $\tau$, the CM  moves with 3-velocity
$v^{\alpha}$
\begin{equation}\label{eq:13}
u^{\alpha}=u^{\lambda}\tau_{\lambda}\left(
\tau^{\alpha}+v^{\alpha}\right),
\end{equation}
\vspace{-1em}
\[
u^{\lambda}\tau_{\lambda}=\left(1-v^{2}\right)^{-1/2}=m/m_{0},
\quad v^{2}\equiv-v^{\alpha}v_{\alpha}
\]
and the rest frame moves with 3-velocity $V^{\alpha}$
\begin{equation}\label{eq:14}
U^{\alpha}=U^{\lambda}\tau_{\lambda}\left(
\tau^{\alpha}+V^{\alpha}\right),
\end{equation}
\vspace{-1em}
\[
U^{\lambda}\tau_{\lambda}=\left(1-V^{2}\right)^{-1/2}=m/M_{0}.
\]
In the rest frame $U$, the CM moves with 3-velocity $w^{\alpha}$
\begin{equation}\label{eq:15}
u^{\alpha}=u^{\lambda}U_{\lambda}\left(
U_{\alpha}+w^{\alpha}\right),
\end{equation}
\vspace{-1em}
\[
u^{\lambda}U_{\lambda}=\left( 1-w^{2}\right)^{-1/2}=M/M_{0}
\]
and the frame $\tau$ moves with 3-velocity $V_{f}^{\alpha}$
\begin{equation}\label{eq:16}
\tau^{\alpha}=\tau^{\lambda}U_{\lambda}\left(
U^{\alpha}+V_{f}{}^{\alpha}\right), \quad V_{f}{}^{2}=V^{2}.
\end{equation}
Finally, $w_{f}{}^{\alpha}$ is the 3-velocity of motion of the
rest frame relative to the CM
\begin{equation}\label{eq:17}
U^{\alpha}=U^{\lambda}u_{\lambda}\left(
u^{\alpha}+w_{f}{}^{\alpha}\right), \quad w_{f}^{2}=w^{2}.
\end{equation}

    We use the units in which $G=c=1$. The signature is
(+ -- -- --); Greek indices run from 0 to 3, and Latin indices run
from 1 to 3; $\varepsilon_{\alpha\beta\mu\nu}$ and
$\epsilon_{\beta\mu\nu}$ are the Levi-Civita 4-tensor and the
spatial Levi-Civita 3-tensor, respectively, in the orthonormalized
basis $\varepsilon^{0123}=\epsilon^{123}=1$. We employ the obvious
simplifying notation according to the rule
\[
\varepsilon_{\alpha\beta\mu\nu}U^{\alpha}u^{\beta}\tau^{\mu}S^{\nu}
= \varepsilon_{Uu\tau S} = \varepsilon^{Uu\tau S}.
\]

    \section{Shift of the CMs\\in flat space-time}

    In flat space-time $R_{\alpha\beta\mu\nu}=0$
the Papapetrou equations describe the conservation of the
4-momentum and of the total angular momentum of a body. In a
coordinate system comoving with the inertial frame $U$
\[
P^{i}=0, \quad \dot{S}^{ik}=0.
\]
Application of a supplementary condition makes it possible to
transport the derivative from the spin tensor in Eq.(\ref{eq:2})
to the projecting vector of the 4-velocity of the frame $\tau$:
\begin{equation}\label{eq:18}
P^{\alpha}u^{\lambda}\tau_{\lambda} = mu^{\alpha} -
\underset{\tau}{S}^{\alpha\beta}\dot{\tau}_{\beta}\,.
\end{equation}
Substitute the expansions $u^{\alpha}$ (\ref{eq:15}) and
$\tau^{\alpha}$ (\ref{eq:16}) into Eq.(\ref{eq:18}) in the
comoving coordinate system $U^{i}=0$:
\begin{equation}\label{eq:19}
M_{0}w^{i} - S^{i}{}_{k}\frac{d}{dt}V_{f}{}^{k} = 0.
\end{equation}
The derivative here and the one in Eq.(\ref{eq:18}) are related by
\[
ds = \sqrt{1-w^{2}}\,dt = \left(M_{0}/{M}\right)\, dt.
\]
The vector form of Eq.(\ref{eq:19}) is
\begin{equation}\label{eq:20}
\bm{w}=\frac{\bm{S}}{M_{0}}\times\frac{d\bm{V}_{f}}{dt}
\end{equation}
Putting $\bm{w}=d\bm{r}/dt$, we obtain the shift vector $\bm{r}$
extended from the intrinsic CM to the CM in the frame $\tau$,
\begin{equation}\label{eq:21}
\bm{r}=\frac{\bm{S}\times\bm{V}_{f}}{M_{0}}\,.
\end{equation}
The CMs defined in a set of inertial frames form a disc \cite{17}
of radius $r_{\text{max}}=S/M_{0}$.

    If we put $\tau=u$, then the relation (\ref{eq:18}) under the Pirani
condition
\begin{equation}\label{eq:22}
P^{\alpha}=Mu^{\alpha} -
\underset{u}{S}^{\alpha\beta}\dot{u}_{\beta}
\end{equation}
leads to Eqs.(\ref{eq:20}) and (\ref{eq:21}), where the
$\bm{V}_{f}$ must be replaced by $\bm{w}$
\begin{equation}\label{eq:23}
\bm{w}=\frac{\bm{S}}{M_{0}}\times\frac{d\bm{w}}{dt}, \quad
\bm{r}=\frac{\bm{S}}{M_{0}}\times\frac{d\bm{r}}{dt}\,.
\end{equation}
These are equations of circular motion of radius $r=wS/M_{0}$ with
angular velocity $M_{0}/S$ opposite to the vector $\bm{S}$
\cite{18}. The Weyssenhoff-Raabe motion (\ref{eq:23}) reflects the
fact that the Papapetrou equations under the Pirani condition turn
out to be of the third order in the derivatives of the
coordinates:
\begin{equation}\label{eq:24}
\dot{P}^{\alpha}=M\dot{u}^{\alpha} -
\underset{u}{S}^{\alpha\beta}\ddot{u}_{\beta}\,.
\end{equation}
The  general relationship between $\dot{P}^{\alpha}$ and
$\dot{u}^{\alpha}$ without supplementary conditions includes the
second derivative of the spin tensor,
\[
\dot{P}^{\alpha}=M\dot{u}^{\alpha}+\ddot{S}^{\alpha\beta}u_{\beta}\,.
\]

    It should be noted that the shift (\ref{eq:21}) and (\ref{eq:23})
can be obtained directly from the supplementary conditions. The
spin tensors $S_{U}^{\alpha\beta}$  and $S^{\alpha\beta}$ are
connected by the relation
\[
\underset{U}{S}^{\alpha\beta}=S^{\alpha\beta} +
r^{\alpha}P^{\beta}-r^{\beta}P^{\alpha}
\]
which follows from Eq.(\ref{eq:3}) at
$\zeta_{U}^{\alpha}=\zeta+r^{\alpha}$. The condition (\ref{eq:6})
in the coordinate system $P^{i}=0$ leads to the M\o ller shift
(\ref{eq:21})
\[
M_{0}r^{k}=S^{k}{}_{i}V_{f}{}^{i},
\]
while the condition (\ref{eq:9}) leads to the Weyssenhoff-Raabe
motion (\ref{eq:23})
\[
M_{0}r^{k}=S^{k}{}_{i}w^{i}.
 \]
Then the Papapetrou equations are satisfied automatically.

    \section{The Corinaldesi supplementary condition}

    From the dual spin tensor
\[
\overset{*}{S}{}^{\alpha\beta} =
\frac{1}{2}\,\varepsilon^{\alpha\beta}{}_{\mu\nu}\,S^{\mu\nu}
\]
one composes the spin vector
\begin{equation}\label{eq:25}
\underset{\tau}{S}^{\alpha}=\tau_{\alpha}\overset{*}{S}{}^{\alpha\beta}
= \frac{1}{2}\,\varepsilon^{\tau\beta}{}_{\mu\nu}\,S^{\mu\nu} =
\frac{1}{2}\,\underset{\tau}{\epsilon}^{\beta}{}_{\mu\nu}\,S^{\mu\nu}\,.
\end{equation}
If the spin tensor satisfies the Colinaldesi condition
(\ref{eq:6}), then
\begin{equation}\label{eq:26}
\underset{\tau}{S}^{\alpha\beta}=
-\varepsilon^{\alpha\beta}{}_{\tau S}, \quad
\overset{*}{\underset{\tau}{S}}{}^{\alpha\beta}=
2\tau^{[\alpha}\underset{\tau}{S}^{\beta]}.
\end{equation}
Using Eq.(\ref{eq:26}), Eq.(\ref{eq:18}) can be rewritten as
\begin{equation}\label{eq:27}
U^{\alpha}u^{\lambda}\tau_{\lambda} =
u^{\alpha}U^{\lambda}\tau_{\lambda} +
M_{0}^{-1}\varepsilon^{\alpha}{}_{\dot{\tau}\tau S}\,.
\end{equation}
The CM and the reference frame $U$ move with the relative velocity
(Eqs.(\ref{eq:13}), (\ref{eq:14}))
\[
V^{\alpha}-v^{\alpha} = M_{0}^{-1}\sqrt{\left(
1-v^{2}\right)\left(1-V^{2}\right)}\,
\varepsilon^{\alpha}{}_{\dot{\tau}\tau S}\,.
\]

    In the momentum transfer equation (\ref{eq:1}) we use
the spin vector (\ref{eq:26}) and the dual Riemann tensor (A.1):
\begin{equation}\label{eq:28}
\dot{P}_{\alpha}=R^{*}_{\alpha u\tau S} =
\frac{m}{m_{0}}\left(R^{*}_{\alpha\tau\tau S} + R^{*}_{\alpha
v\tau S}\right)\,,
\end{equation}
\begin{equation}\label{eq:29}
R^{*}_{\alpha\tau\tau S} = - \underset{\tau}{B}{}_{\alpha S}\, ,
\quad -\underset{\tau}{h}{}^{\beta}_{\alpha}R^{*}_{\beta v\tau S}
= \underset{\tau}{\epsilon}{}_{v\gamma\alpha}
\underset{\tau}{E}{}^{\gamma}_{S}\,.
\end{equation}
Here
\[
\underset{\tau}{E}{}_{\alpha\beta} = R_{\alpha\tau\beta\tau}\, ,
\quad \underset{\tau}{B}{}_{\alpha\beta} =
R^{*}_{\alpha\tau\beta\tau}
\]
are the "electric" and "magnetic" parts of the Riemann tensor
(A.2) - (A.3) in the frame $\tau$, and
\[
\underset{\tau}{h}{}_{\alpha\beta}=\tau_{\alpha}\tau_{\beta}-g_{\alpha\beta}
\]
is the metric tensor of the local 3-space orthogonal to
$\tau^{\alpha}$.

    The transport equation (\ref{eq:2}) for the spin vector
(\ref{eq:25})--(\ref{eq:26}) become
\begin{equation}\label{eq:30}
\underset{\tau}{\dot{S}}^{\alpha} +
\tau^{\alpha}\underset{\tau}{S}^{\lambda}\dot{\tau}_{\lambda} =
\underset{\tau}{\epsilon}{}^{\alpha}{}_{\mu\nu}\,P^{\mu}u^{\nu}\,.
\end{equation}
The right-hand side can be expressed in terms of $v^{\alpha}$ and
$V^{\alpha}$ (\ref{eq:13})--(\ref{eq:14}):
\[
\underset{\tau}{\epsilon}{}^{\alpha}{}_{\mu\nu}\,P^{\mu}u^{\nu} =
\frac{m^{2}}{m_{0}}\,\,\underset{\tau}{\epsilon}{}^{\alpha}{}_{\mu\nu}\,V^{\mu}v^{\nu}\,.
\]
The spin vector transport operator is
\[
\underset{\tau}{\dot{S}}^{\alpha} +
\tau^{\alpha}\underset{\tau}{S}^{\lambda}\dot{\tau}_{\lambda} = -
\underset{\tau}{h}{}^{\alpha}_{\beta}\,\underset{\tau}{\dot{S}}^{\beta}.
\]
The  masses (\ref{eq:10})--(\ref{eq:12}) are not conserved:
\begin{equation}\label{eq:31}
\dot{m}=R^{*}_{\tau u \tau S}+P^{\alpha}\dot{\tau}_{\alpha}\,,
\quad \dot{M}_{0}=R^{*}_{Uu\tau S} \,,
\end{equation}
\begin{equation}\label{eq:32}
\dot{M}u^{\lambda}\tau_{\lambda} =
\varepsilon_{\dot{u}\dot{\tau}\tau S}\,, \quad m_{0} = M +
\varepsilon_{\dot{\tau}v\tau S}\,.
\end{equation}
The quantity $\dot{\tau}_{\alpha} = u^{\lambda}\tau_{\alpha ;
\lambda}$ is
\begin{equation}\label{eq:33}
\dot{\tau}_{\alpha} = (1-v^{2})^{-1/2}\left[a_{\alpha} +
v^{\beta}\left(A_{\alpha\beta} - D_{\alpha\beta}\right)\right],
\end{equation}
where \cite{19}
\[
a_{\alpha}=\tau^{\lambda}\tau_{\alpha ; \lambda},\,
A_{\alpha\beta}=h^{\mu}_{\alpha}h^{\nu}_{\beta}\tau_{[\mu ;
\nu]},\, D_{\alpha\beta} =
-h^{\mu}_{\alpha}h^{\nu}_{\beta}\tau_{(\mu ; \nu)}
\]
are, the acceleration vector, the angular velocity tensor and the
rate-of-strain tensor of the frame $\tau$, respectively. Taking
into account Eq.(\ref{eq:33}), the first equation (\ref{eq:31})
can be written as
\begin{equation}\label{eq:34}
\frac{dm}{d\tau} = R^{*}_{\tau v \tau S } +
m\left(a_{\alpha}v^{\alpha} - D_{\mu\nu}v^{\mu}v^{\nu}\right).
\end{equation}

    If space-time possesses the Killing vector
$\xi_{\mu}$, do that $\xi_{\mu ;\nu}+\xi_{\nu ;\mu}=0$, then the
scalar
\begin{equation}\label{eq:35}
K=P^{\mu}\xi_{\mu} - \frac{1}{2}S^{\mu\nu}\xi_{\mu;\nu}
\end{equation}
is the integral of motion, $\dot{K}=0$ \cite{13}. For the
conservation of $K$, no supplementary conditions are required. Let
us direct the $\tau^{\mu}$ along the time-like Killing vector:
\[
\tau^{\mu}=\left(\xi^{\lambda}\xi_{\lambda}\right)^{-1/2}\xi^{\mu}
\]
Then, under the Corinaldesi condition, the quantity
\begin{equation}\label{eq:36}
m_{\xi}=\sqrt{\xi^{\lambda}\xi_{\lambda}}\left(m+S^{\alpha}A_{\alpha}\right)
\end{equation}
is conserved, where
$A^{\alpha}=\epsilon^{\alpha}{}_{\mu\nu}A^{\mu\nu}$ is the vector
of the angular velocity of the frame. This quantity $m_{\xi}$ can
be named as a Killingian mass.

    If we put $P^{\alpha}=Mu^{\alpha}$ (\ref{eq:4}) and associate
the spin tensor with a vector according to the rule
(\ref{eq:25})--(\ref{eq:26}), then the parallel transport of the
spin tensor (\ref{eq:5}) leads to the transport of the vector
$S^{\alpha}$
\[
\underset{\tau}{h}{}^{\alpha}_{\beta}\,\dot{S}^{\beta} = 0.
\]
In this case the length of the spin vector and the mass $M=M_{0}$
are conserved.

     \section{The Pirani supplementary \\ condition}

    Similarly to Eq.(\ref{eq:25}) or simply putting
$\tau^{\alpha}=u^{\alpha}$, we associate the spin tensor with the
vector
\vspace{-0.2em}%
\begin{equation}\label{eq:37}
\underset{u}{S}^{\beta}=u_{\alpha}\overset{*}{S}{}^{\alpha\beta},
\quad \underset{u}{S}^{\beta}u_{\beta}=0.
\vspace{-0.2em}%
\end{equation}
Using the Pirani condition (\ref{eq:9}), we can express the spin
tensor in terms of the vector (\ref{eq:37})
\vspace{-0.2em}%
\begin{equation}\label{eq:38}
\underset{u}{S}^{\alpha\beta}= -\varepsilon^{\alpha\beta}{}_{uS},
\quad \overset{*}{\underset{u}{S}}{}^{\alpha\beta}=
2u^{[\alpha}\underset{u}{S}^{\beta]}.
\vspace{-0.2em}%
\end{equation}
Eq.(\ref{eq:27}), (\ref{eq:22}) appears in the form (\ref{eq:17}):
\begin{equation}\label{eq:39}
U^{\alpha}=\frac{M}{M_{0}}\left(u^{\alpha}+w_{f}{}^{\alpha}\right),
\quad
w_{f}{}^{\alpha}=\frac{1}{M}\,\varepsilon^{\alpha}{}_{\dot{u}
uS}\,.
\end{equation}
It is noteworthy that under the Pirani condition the projection of
the spin vector onto the 4-momentum is zero
\vspace{-0.2em}%
\begin{equation}\label{eq:40}
\underset{u}{S}^{\alpha}U_{\alpha}=0
\vspace{-0.2em}%
\end{equation}
which is evident from Eq.(\ref{eq:39}).

    The momentum transfer equation (\ref{eq:1}), (\ref{eq:28}) has
a simple appearance
\vspace{-0.2em}%
\begin{equation}\label{eq:41}
\dot{P}^{\alpha}=-\underset{u}{B}{}^{\alpha}_{S}\,,
\vspace{-1em}%
\end{equation}
where
\[
\underset{u}{B}{}_{\alpha\beta}=R^{*}_{\alpha u\beta u}
\]
is the "magnetic" part of the Riemann tensor (A.3) in the frame
$u$. But Eq.(\ref{eq:41}) includes (\ref{eq:24}) the second
derivative $\ddot{u}^{\alpha}$:
\vspace{-0.2em}%
\[
\dot{P}^{\alpha}=M\dot{u}^{\alpha} +
\varepsilon^{\alpha}{}_{\ddot{u} uS}
\vspace{-0.2em}%
\]

    The spin vector under the Pirani condition is transported
according to Fermi-Walker:
\begin{equation}\label{eq:42}
\underset{u}{\dot{S}}^{\alpha} +
u^{\alpha}\underset{u}{S}^{\lambda}\dot{u}_{\lambda}=0.
\end{equation}
In this case, in contrast to the transport (\ref{eq:30}), a length
of the spin vector is conserved,
\vspace{-0.2em}%
\[
\underset{u}{S}{}_{\alpha}\underset{u}{\dot{S}}^{\alpha}=0.
\vspace{-0.2em}%
\]
The mass $M$ (\ref{eq:11}) is also conserved:
\vspace{-0.2em}%
\[
\dot{M}=\varepsilon_{\dot{u}\dot{u} uS}=0.
\vspace{-0.2em}%
\]
The mass $M_{0}$ (\ref{eq:10}) is not conserved:
\vspace{-0.2em}%
\[
\dot{M}_{0}=R^{*}_{UuuS} = M_{0}^{-1}\varepsilon_{\dot{P}\dot{u}
us}.
\vspace{-0.2em}%
\]

    Despite of the relation (\ref{eq:40}) resembling the Dixon
condition, the Pirani condition in curved space-time does not
allow one to single out the intrinsic CM. None of the world lines
of the Pirani CM satisfies the Dixon condition:
\begin{equation}\label{eq:43}
\underset{u}{S}{}^{\alpha\beta}U_{\beta}=-\varepsilon^{\alpha}{}_{UuS}\,,
\quad \dot{u}_{\alpha}\underset{u}{S}{}^{\alpha\beta}P_{\beta} =
M_{0}{}^{2} - M^{2}.
\vspace{-0.2em}%
\end{equation}

    Note that it follows from Eq.(\ref{eq:39}) that
\vspace{-0.2em}%
\[
M\underset{u}{S}^{\alpha}\dot{u}_{\alpha} =
\underset{u}{S}^{\alpha}\dot{P}_{\alpha}
\vspace{-0.2em}%
\]
therefore the Fermi-Walker transport equation with the use of
Eq.(\ref{eq:41}) can be written as
\begin{equation}\label{eq:44}
M\underset{u}{\dot{S}}^{\alpha} =
u^{\alpha}\underset{u}{B}{}_{SS}\,.
\end{equation}

    \section{The Dixon supplementary \\ condition}

    Similarly to Eqs.(\ref{eq:25}) and (\ref{eq:37}),
the spin vector is
\[
\underset{U}{S}^{\beta}=U_{\alpha}\overset{*}{S}{}^{\alpha\beta},
\quad \underset{U}{S}^{\beta}U_{\beta}=0.
\]
Using the Dixon condition (\ref{eq:7})
$S_{U}^{\alpha\beta}U_{\beta}=0$, we obtain the relations
\[
\underset{U}{S}^{\alpha\beta}= -\varepsilon^{\alpha\beta}{}_{US},
\quad \overset{*}{\underset{U}{S}}{}^{\alpha\beta}=
2U^{[\alpha}\underset{U}{S}^{\beta]}.
\]
The equation relating the kinematic and dynamic 4-velocities,
obtained by substituting $\tau^{\alpha}=U^{\alpha}$ into
Eq.(\ref{eq:27}), appears in the form (\ref{eq:15}):
\begin{equation}\label{eq:45}
u^{\alpha}=\frac{M}{M_{0}}\left(U^{\alpha}+w^{\alpha}\right),
\quad w^{\alpha}=\frac{1}{M}\,\varepsilon^{\alpha}{}_{\dot{U}
US}\,
\end{equation}
whence it follows, in particular, that
\begin{equation}\label{eq:46}
\underset{U}{S}^{\alpha}u_{\alpha}=0.
\end{equation}

    The 4-momentum transfer equation (\ref{eq:1}), (\ref{eq:28})
becomes
\begin{equation}\label{eq:47}
\dot{P}_{\alpha}=R^{*}_{\alpha UUS} =
-\frac{M}{M_{0}}\left(\underset{U}{B}{}^{\alpha}_{S} -
\epsilon^{\alpha}{}_{w\beta}\underset{U}{E}{}^{\beta}_{S}\right).
\end{equation}
The spin part of the Papapetrou equations (\ref{eq:2}),
(\ref{eq:30}) describes the Dixon transport
\begin{equation}\label{eq:48}
\underset{U}{\dot{S}}^{\alpha} +
U^{\alpha}\underset{U}{S}^{\lambda}\dot{U}_{\lambda} = 0.
\end{equation}
Using the relation
\[
M\underset{U}{S}^{\alpha}\dot{U}_{\alpha} =
M_{0}\underset{U}{S}^{\alpha}\dot{u}_{\alpha},
\]
which is obtained with the aid of Eqs.(\ref{eq:45}), from the
Dixon transport (\ref{eq:48}) we can isolate the Fermi-Walker
transport (\ref{eq:42}):
\[
\underset{U}{S}^{\alpha} +
u^{\alpha}\underset{U}{S}^{\lambda}\dot{u}_{\lambda} =
M^{-1}\varepsilon^{\alpha}{}_{u\dot{U} S}\,
\underset{U}{S}^{\lambda}\dot{u}_{\lambda}\,.
\]
Just as the Fermi-Walker transport, the Dixon transport conserves
the length of the spin vector,
\[
\underset{U}{S}{}_{\alpha}\underset{U}{\dot{S}}^{\alpha}=0.
\]
The mass $M_{0}$ (\ref{eq:10}) is also conserved
\[
\dot{M}_{0}=R^{*}_{UuUS}=(M_{0}/M)R^{*}_{uuUS} +
M^{-1}\dot{P}^{\lambda}\,\varepsilon_{\lambda\dot{U}US}=0.
\]
The mass $M$ (\ref{eq:11}) under the Dixon condition is not
conserved
\[
\dot{M}=(M_{0}/M)\,\varepsilon_{\dot{u}\dot{U}uS}\,.
\]

    The velocity of motion  $w^{\alpha}$ (\ref{eq:45}) of the intrinsic
CM in the rest frame $U$ is not zero:
\begin{equation}\label{eq:49}
w^{\alpha}=\left(M_{0}{}^{2} + \underset{U}{E}{}_{SS}\right)^{-1}
\epsilon^{\alpha S\lambda} \underset{U}{B}{}_{\lambda S}\,.
\end{equation}
On the world line of the Dixon CM, the Pirani condition is not
fulfilled:
\begin{multline}\label{eq:50}
\underset{U}{S}{}^{\alpha\beta}u_{\beta}=
-\varepsilon^{\alpha}{}_{uUS}=\\
=\frac{M}{M_{0}}\left( M_{0}{}^{2} +
\underset{U}{E}{}_{SS}\right)^{-1}\!
\left(\underset{U}{S}{}^{\lambda}
\underset{U}{S}{}_{\lambda}\underset{U}{B}{}^{\alpha}_{S} -
\underset{U}{S}{}^{\alpha}\underset{U}{B}{}_{SS}\right),
\end{multline}
\begin{equation}\label{eq:51}
\quad \dot{P}_{\alpha}\underset{U}{S}{}^{\alpha\beta}u_{\beta} =
M^{2} - M_{0}{}^{2} .
\end{equation}
The Dixon and Pirani supplementary conditions single out different
world lines.

    \section{Vertical spin in a circular\\ orbit in a static
       axial field}

    As an example, we consider the motion of a body with spin
orthogonal to the plane of an orbit of constant radius $u^{1}=0$
with the Corinaldesi, Pirani and Dixon supplementary conditions.
Such a motion is possible in the Schwarzschild metric and for an
equatorial orbit, in any axial-symmetric stationary metric.
Confining ourselves to static space-time, we ignore the spin-spin
interaction.

    For simplicity of representation, we use the orthonormal
basis comoving with a rigid ($D_{\alpha\beta}=0$) nonrotating
($A_{\alpha\beta}=0$) reference frame in which the "magnetic" part
of the Riemann tensor (A.3) of static space-time is zero. For
example, in the Schwarzschild metric in a frame at rest with
respect to the curvature coordinates
\begin{equation}\label{eq:52}
E^{i}_{k}=\frac{\mathcal{M}}{r^{3}}\left( \begin{array}{ccc}
-2&0&0\\0&1&0\\0&0&1 \end{array}
  \right), \quad B^{i}_{k}=0,
\end{equation}
where $\mathcal{M}$ is the mass of a source and $r$ is the radial
curvature coordinate.

    The radial Papapetrou equation, combined with the 4-velocity
normalization condition, fully determines the motion of a body.
Eqs.(\ref{eq:27}), (\ref{eq:39}) ore (\ref{eq:45}) yield a
relationship between the kinematic and dynamic 4-velocities. The
quantities $M_{0}$, $M$, $m$ and $m_{0}$
(\ref{eq:10})--(\ref{eq:12}) are conserved in the case of such a
motion. The quantity $u^{3}S^{2}$ is negative for the spin vector
parallel to the vector of the angular velocity of revolution.

    The Corinaldesi condition in the basis used signifies simply
$S^{0i}=0$, and the set of defining equations appears as
\begin{equation}\label{eq:53}
m_{0}\dot{u}^{1} =
-\left(E^{2}_{2}+\gamma^{1}{}_{33}a^{1}\right)u^{3}S^{2},
\end{equation}
\begin{equation}\label{eq:54}
P^{3} =m_{0}u^{3}+a^{1}S^{2}, \quad P^{0}=m_{0}u^{0},
\end{equation}
where
\[
\dot{u}^{1}=\gamma^{1}{}_{00}\left(u^{0}\right)^{2} +
\gamma^{1}{}_{33}\left(u^{3}\right)^{2}, \quad
\left(u^{0}\right)^{2} - \left(u^{3}\right)^{2} = 1.
\]
The Pirani condition singles out another word line with the
constant radial coordinate
\begin{multline}\label{eq:55}
\dot{u}^{1}\left[M+\left(\gamma^{0}{}_{10} -
\gamma^{3}{}_{13}\right)u^{0}u^{3}S^{2}\right] =\\
=\left(E^{1}_{1}-E^{2}_{2}\right)u^{0}u^{3}S^{2},
\end{multline}
\begin{equation}\label{eq:56}
P^{3}=Mu^{3}+\dot{u}^{1}u^{0}S^{2}, \quad
P^{0}=Mu^{0}+\dot{u}^{1}u^{3}S^{2}.
\end{equation}

    The quantity $u^{3}/u^{0}$ is the velocity of revolution according
to the clock of the frame (\ref{eq:52}), which is obtained
difference from Eqs.(\ref{eq:53}) and (\ref{eq:55}) even in the
post-Newtonian approximation. The formula for the angular velocity
of revolution also turns out be different. Namely, the Pirani
condition gives an angular velocity $\omega_{u}$ which differs
from that of a nonrotating body $\Omega=\sqrt{\mathcal{M}/r^{3}}$
\cite{20}:
\begin{equation}\label{eq:57}
\omega_{u}=\Omega\left(1-\frac{3}{2}\,\Omega S\right)
 \end{equation}
where $S\equiv S^{2}$ whereas under the Corinaldesi condition the
CM revolves with the angular velocity of a nonrotating body,
\[
\omega_{\tau}=\Omega
\]
(the right side of Eq.(\ref{eq:53}) becomes zero in the
approximation linear with respect to $\mathcal{M}$).

    The question is whether it means that the Pirani and
Corinaldesi CMs drift and can be found at any mutual distance? The
point is that in accordance with Eq.(\ref{eq:21}) the Corinaldesi
and Pirani CMs are shifted radially
\begin{equation}\label{eq:58}
r_{\tau}=r_{u}\left(1+\Omega S\right).
\end{equation}
Therefore the angular velocities $\omega_{u}$ and $\omega_{\tau}$
are actually the same
\begin{equation}\label{eq:59}
\omega_{u} = \sqrt{\frac{\mathcal{M}}{r_{u}^{3}}}
\left(1-\frac{3}{2}\,\Omega S\right) =
\sqrt{\frac{\mathcal{M}}{r_{\tau}^{3}}} = \omega_{\tau},
\end{equation}
the CMs are shifted and do not drift. The Pirani CM accordingly
drifts and rests relative to different nonrotating bodies with the
orbital radii $r_{u}$ and $r_{\tau}$. The revolution velocities
$v$ of different CMs of the same body are different
\[
\left(u^{3}/u^{0}\right)_{\tau}=v_{\tau}=v_{u}\left(1+\Omega
S\right).
\]
The dynamic velocity under the Corinaldesi condition according to
Eq.(\ref{eq:54}) is the same as that under the Pirani condition
\[
\left(P^{3}/P^{0}\right)_{u}=v_{u}=\left(P^{3}/P^{0}\right)_{\tau}\,.
\]

    The shift (\ref{eq:21}) is written with respect to the intrinsic
CM determined by the Dixon condition. In the post-Newtonian
approximation, the Pirani and Dixon CMs in a circular orbit
coincide, as well as the velocities of revolution. The exact world
lines of the CMs according to Dixon and Pirani are different,
which is immediately evident from Eq.(\ref{eq:50}), taking into
account that
\[
\underset{U}{B}{}^{1}_{S}=\left(E^{1}_{1}-E^{2}_{2}\right)U^{0}U^{3}S^{2},
\quad S^{1}=0.
\]

    The motion of a body with horizontal spin in the post-
Newtonian approximation under the Pirani (Dixon) condition has
been investigated in Ref.\cite{21}. In the next section, we
consider the general case of the spin-orbit interaction in the
post-Newtonian approximation under different supplementary
conditions.

    \section{Spin-orbital force in the post-Newtonian
    approximation}

    The leading post-Newtonian approximation for the spin-orbital
force denotes a linear approximation with respect to the orbital
motion velocity $v$, spin $S$ and mass $\mathcal{M}$ of the
source. The masses (\ref{eq:10})--(\ref{eq:12}) are
$M=M_{0}=m_{0}=m$. The leading approximation of the spin-orbital
force $F_{S}$ is
\begin{equation}\label{eq:60}
\frac{F_{S}}{F_{N}}\sim\frac{vS}{mr} \sim
\frac{\mathcal{M}}{r}\frac{S}{mvr},
\end{equation}
where $F_{N}=\mathcal{M}m/r^{2}$. In the frame resting relative to
the source of a static field $B_{ik}=0$ (\ref{eq:52}). In the
absence of rotation of the spatial axes, according to Eq.(A.7) we
have
\[
\underset{u}{B}^{ij}=-2\epsilon^{vk(i}E^{j)}_{k}.
\]
The spin-orbital force under the Dixon and Pirani conditions in
the leading approximation is the same:
\begin{equation}\label{eq:61}
F_{SU}^{i}=F_{Su}^{i}=2\epsilon^{vk(i}E^{S)}_{k}.
\end{equation}
The spin-orbital force under the Corinaldesi condition differs
from (\ref{eq:61}) in the leading approximation
\begin{equation}\label{eq:62}
F_{S\tau}^{i}=-2\epsilon^{ik(S}E^{v)}_{k}\,.
\end{equation}
Let us write a general expression for the force (\ref{eq:61}),
(\ref{eq:62}):
\begin{equation}\label{eq:63}
F_{S}^{i} =
F_{Su}^{i}+\sigma\left(F_{S\tau}^{i}-F_{Su}^{i}\right),
\end{equation}
\vspace{-5ex}
\begin{multline}\label{eq:64}
\bm{F}_{S}=3\frac{\mathcal{M}}{r^{3}}[\bm{S}\times\bm{v}
    +(2 - \sigma)\,\hat{\bm{r}}\,
(\bm{S}(\hat{\bm{r}}\times\bm{v}))- \\
 -(1 + \sigma)(\bm{v}\hat{\bm{r}})(\bm{S}\times\hat{\bm{r}})]
\end{multline}
where $\hat{\bm{r}}=\bm{r}/r$. The Dixon-Pirani condition
corresponds to $\sigma=0$, the Corinaldesi condition to
$\sigma=1$, and $\sigma=1/2$ leads to the results of Fock
\cite{22} and Refs.\cite{14, 15}. The Corinaldesi, Dixon-Pirani,
Fock and \cite{14, 15} supplementary conditions in the
approximation used can be written as
\vspace{-0.2em}%
\[
S^{0i}=\left(\lambda-1\right)v_{k}S^{ki}
\vspace{-0.2em}%
\]
The representing points move differently under different
supplementary conditions, but all differences reduce to a shift of
the CMs in accordance with Eq.(\ref{eq:21})
\begin{equation}\label{eq:65}
\bm{r}=\bm{r}_{u}+\sigma\bm{v}\times\bm{S}/m.
\end{equation}
The $\sigma$ dependencies on the left- and right-hand sides of the
equation $m\dot{u}^{i}=F_{s}^{i}$ are mutually cancelled. In fact,
the expressions following from Eq.(\ref{eq:65}) for the body
acceleration and for the Newtonian attraction force
\[
m\frac{dv^{i}}{d\tau} =
m\frac{dv_{u}^{i}}{d\tau}+\sigma\epsilon^{Ski}E^{v}_{k},
\]
\[
m\mathcal{M}\frac{\hat{r}^{i}}{r^{2}} =
m\mathcal{M}\frac{\hat{r}^{i}_{u}}{r^{2}_{u}} +
\sigma\epsilon^{Skv}E^{i}_{k}
\]
indicate that the quantity
\begin{equation}\label{eq:66}
m\left(\frac{dv^{i}}{d\tau} +
\mathcal{M}\frac{\hat{r}^{i}}{r^{2}}\right) - F_{S}^{i}
\end{equation}
is independent of $\sigma$.

    Compare Eq.(\ref{eq:64}) with a force exerted on a nonrotating body
$\mathcal{M}$ moving with velocity $-\bm{v}$ in the field of
rotating mass $\bm{S}$ (gravitomagnetic Coriolis force)
\begin{equation}\label{eq:67}
\bm{F}_{\mathcal{M}} = -2\mathcal{M}\bm{v}\times\frac{\bm{S} -
3\hat{\bm{r}}\left(\bm{S}\hat{\bm{r}}\right)}{r^{3}}.
\end{equation}
It can be seen that Eq.(\ref{eq:64}) can not be reduced to the
form (\ref{eq:67}). The case is different with electrodynamics. A
force $\bm{F}_{J}$, exerted on a magnetic dipole $\bm{J}$
\cite{23} during its motion with velocity $\bm{v}$ in the field of
charge $Q$ in the approximation linear with respect to $v$
\cite{24}, is
\begin{equation}\label{eq:68}
\bm{F}_{J} = \left(\bm{J}\nabla\right)\bm{H} =
-\frac{Q}{r^{3}}\bm{v}\times\left(\bm{J} -
3\hat{\bm{r}}\left(\bm{J}\hat{\bm{r}}\right)\right),
 \end{equation}
where $\bm{H}$ is a magnetic field in the frame comoving with the
dipole, $\bm{H}=-\bm{v}\times\bm{E}$,
$\bm{E}=Q\hat{\bm{r}}/r^{3}$. The Lorentz force $\bm{F}_{Q}$,
exerted on the charge $Q$ in the magnetic field of the dipole
$\bm{J}$, is
\begin{equation}\label{eq:69}
\bm{F}_{Q} = Q\bm{v}\times\frac{\bm{J} -
3\hat{\bm{r}}\left(\bm{J}\hat{\bm{r}}\right)}{r^{3}} =
-\bm{F}_{J}\,.
\end{equation}
In expressions (\ref{eq:68}) and (\ref{eq:69}), $v$ is a relative
velocity of the dipole $J$ and the charge $Q$.

    Let us write out the result, corresponding to Eqs.(\ref{eq:68}) and
(\ref{eq:69}), of the two-body problem in general relativity
\cite{25}. The first body has spin $\bm{S}$ and velocity
$\bm{v}_{1}$; the mass of the second body is $\mathcal{M}$ and its
velocity is $\bm{v}_{2}$; $\bm{v}=\bm{v}_{1}-\bm{v}_{2}$,
$\bm{r}=\bm{r}_{1}-\bm{r}_{2}$; frame is arbitrary.
\begin{multline}\label{eq:70}
\bm{F}_{1} = \frac{\mathcal{M}}{r^{3}}[3\bm{S}\times\bm{v}_{1} -
(3+\sigma)\bm{S}\times\bm{v}_{2} +\\
+(6-3\sigma)\hat{\bm{r}}\left(\bm{S}(\hat{\bm{r}}\times\bm{v}_{1})\right)
- 6\hat{\bm{r}}\left(\bm{S}(\hat{\bm{r}}\times\bm{v}_{2})\right)-\\
-(3+3\sigma)(\bm{v}\hat{\bm{r}})(\bm{S}\times\hat{\bm{r}})]\,,
\end{multline}
\vspace{-6ex}
\begin{multline}\label{eq:71}
\bm{F}_{2} =
-\frac{\mathcal{M}}{r^{3}}[(4-\sigma)\bm{S}\times\bm{v}_{1} -
4\bm{S}\times\bm{v}_{2} +\\
+(6-3\sigma)\hat{\bm{r}}\left(\bm{S}(\hat{\bm{r}}\times\bm{v}_{1})\right)
-6\hat{\bm{r}}\left(\bm{S}(\hat{\bm{r}}\times\bm{v}_{2})\right)-\\
-6(\bm{v}\hat{\bm{r}})(\bm{S}\times\hat{\bm{r}})]\,.
\end{multline}
To compare $\bm{F}_{2}$ (\ref{eq:71}) and $\bm{F}_{\mathcal{M}}$
(\ref{eq:67}), one should keep in mind the vector identity
\begin{multline*}
2\bm{S}\times\bm{v} +
3\hat{\bm{r}}\left(\bm{S}(\hat{\bm{r}}\times\bm{v})\right) -
3(\bm{v}\hat{\bm{r}})(\bm{S}\times\hat{\bm{r}})=\\
=\bm{v}\times\left(\bm{S} -
3\hat{\bm{r}}\left(\bm{S}\hat{\bm{r}}\right)\right).
\end{multline*}

For $\sigma=0$ (Pirani), only the relative velocity
$\bm{v}_{1}-\bm{v}_{2}$ occurs in the expressions for the forces
$F_{1}$ and $F_{2}$,  whereas for the fulfilment of Newton's third
law $\bm{F}_{2}=-\bm{F}_{1}$ the condition $\sigma=1$
(Corinaldesi) is required. It should be stressed that the
quantities (\ref{eq:66})
\[
m\left(\frac{d\bm{v}_{1}}{d\tau} +
\mathcal{M}\frac{\hat{\bm{r}}}{r^{2}}\right) - \bm{F}_{1}, \quad
\mathcal{M}\left(\frac{d\bm{v}_{2}}{d\tau} +
m\frac{\hat{\bm{r}}}{r^{2}}\right) - \bm{F}_{2}
\]
are independent of $\sigma$, and the equations
\[
m\dot{u}_{1}^{i}=F_{1}^{i}, \quad
\mathcal{M}\dot{u}_{2}^{i}=F_{2}^{i}
\]
lead to motions of the bodies $m$ and $\mathcal{M}$, which are
independent of supplementary conditions. However, the $\bm{F}_{1}$
and $\bm{F}_{2}$ , $\bm{F}_{S}$ and $\bm{F}_{\mathcal{M}}$ at any
$\sigma$ do not possesses electrodynamic symmetry
(\ref{eq:68})--(\ref{eq:69}), which indicates that it is
impossible to satisfy the set of the third Newton law and the
relativity of motions in the theory of gravity.

    \section*{Appendix.\\"Electric" and "magnetic" parts of the Riemann
    tensor}

    Associate the Riemann tensor $R_{\alpha\beta\mu\nu}$ with dual
tensors
\[
R^{*}_{\alpha\beta\mu\nu} = (1/2)R_{\alpha\beta\gamma\delta}
\varepsilon^{\gamma\delta}{}_{\mu\nu}, \eqno (A.1)
\]
\[
{}^{*}\!R_{\alpha\beta\mu\nu} =
(1/2)\,\varepsilon_{\alpha\beta}{}^{\sigma\lambda}
R_{\sigma\lambda\mu\nu} = R^{*}_{\mu\nu\alpha\beta},
\]
\[
{}^{*}\!R^{*}_{\alpha\mu\nu} =
(1/4)\,\varepsilon_{\alpha\beta}{}^{\sigma\lambda}
R_{\sigma\lambda\gamma\delta}\varepsilon^{\gamma\delta}{}_{\mu\nu}.
\]
In the orthonormal basis $\epsilon_{0ijk}=\epsilon_{ijk}$ the
Riemann tensor can be presented in the form of a $6\times6$ matrix
\cite{10}
\[
R^{M}{}_{N}= \left(
\begin{array}{cc}
R^{i0}{}_{k0}&R^{*}{}^{i0}{}_{k0}\\
-{}^{*}\!R^{i0}{}_{k0}&{}^{*}\!R^{*}{}^{i0}{}_{k0}
\end{array}
\right) =
 \left(
\begin{array}{cc}
E&B\\-B^{T}&C
\end{array}
\right)
\]
where $M$ and $N$ are collective indices (10, 20, 30, 23, 31, 12),
$E$ and $B$ is  "electric" and "magnetic" $3\times3$ matrices
\[
E^{i}_{k} = R^{i0}{}_{k0}, \quad E=E^{T}, \eqno (A.2)
\]
\[
B^{i}_{k} = R^{*}{}^{i0}{}_{k0} =
-\frac{1}{2}R^{i0}{}_{mn}\epsilon^{mn}{}_{k}, \quad B^{i}_{i}=0,
\eqno (A.3)
\]
\[
C^{i}_{k}={}^{*}\!R^{*}{}^{i0}{}_{k0}, \quad C=C^{T}.
\]
In empty space-time, Riemann tensor is split into "electric" and
"magnetic" parts only:
\[
R^{M}{}_{N}= \left(
\begin{array}{cc}
E&B\\-B&E
\end{array}
\right),
\]
\[
C=E, \quad E^{i}_{i}=0, \quad B=B^{T}.
\]

In passing to a new basis $e'^{\nu}$,
$e^{\mu}=L^{\mu}{}_{\nu}e'^{\nu}$, the "electric" and "magnetic"
matrices are transformed according to the law \cite{26}
\begin{multline*}
E'_{ij}=4E_{kl}L^{[k}{}_{i}u^{0]}L^{[l}{}_{j}u^{0]}-\\
       -E_{pq}\epsilon^{p}{}_{km}\epsilon^{q}{}_{ln}
             L^{k}{}_{i} u^{m} L^{l}{}_{j} u^{n}+ \\
 +4B_{km}\epsilon^{m}{}_{ln}
 L^{[k}{}_{(i}u^{0]}L^{[l}{}_{j)}u^{n]}\,, \qquad (A.4)
  \end{multline*}
\vspace{-4ex}
\begin{multline*}
B'_{ij}=4B_{kl}L^{[k}{}_{i}u^{0]}L^{[l}{}_{j}u^{0]}-\\
       -B_{pq}\epsilon^{p}{}_{km}\epsilon^{q}{}_{ln}
             L^{k}{}_{i} u^{m} L^{l}{}_{j} u^{n}- \\
 -4E_{km}\epsilon^{m}{}_{ln}
 L^{[k}{}_{(i}u^{0]}L^{[l}{}_{j)}u^{n]}\,. \qquad (A.5)
  \end{multline*}
where $u^{\mu}=L^{\mu}{}_{0}$ are components of the 4-velocity in
the basis $e^{\mu}$. If there is no rotation of the spatial axes,
in the approximation linear with respect to the velocity
$v^{i}=u^{i}$
\[
 E'_{ij}=E_{ij}+2v^{l}\epsilon_{lk(i}B^{k}_{j)}\,, \eqno
(A.6)
\]
\[
B'_{ij}=B_{ij}-2v^{l}\epsilon_{lk(i}E^{k}_{j)}\,. \eqno (A.7)
\]

\end{document}